# Spin Hall effects in mesoscopic Pt films with high resistivity

Chuan Qin,[1] Yongming Luo,[2,3] Chao Zhou,[2] Yunjiao Cai,[1] Mengwen Jia,[2] Shuhan Chen,[1] Yizheng Wu,[2] and Yi Ji[1*]

1. Department of Physics and Astronomy, University of Delaware, Newark, Delaware 19716, U.S.A.

2. Department of Physics, State Key Laboratory of Surface Physics, and Collaborative Innovation Center of Advanced Microstructures, Fudan University, Shanghai 200433, P.R. China

3. Center for Integrated Spintronic Devices, Hangzhou Dianzi University, Hangzhou, Zhejiang 310018, P.R. China

*Email: yji@udel.edu

**Abstract:** The energy efficiency of the spin Hall effects (SHE) can be enhanced if the electrical conductivity is decreased without sacrificing the spin Hall conductivity. The resistivity of Pt films can be increased to 150 – 300 μΩ•cm by mesoscopic lateral confinement, thereby decreasing the conductivity. The SHE and inverse spin Hall effects (ISHE) in these mesoscopic Pt films are explored at 10 K by using the nonlocal spin injection/detection method. All relevant physical quantities are determined in-situ on the same substrate, and a quantitative approach is developed to characterize all processes effectively. Extensive measurements with various Pt thickness values reveal an upper limit for the Pt spin diffusion length: $\lambda_{pt}$ ≤ 0.8 nm. The average product of $\lambda_{pt}$ and the Pt spin Hall angle $\alpha_H$ is substantial: $\alpha_H \lambda_{pt}$ = (0.142 ± 0.040) nm for 4 nm thick Pt, though a gradual decrease is observed at larger Pt thickness. The results suggest enhanced spin Hall effects in resistive mesoscopic Pt films.

**I, Introduction**

Spin Hall effects (SHE) and inverse spin Hall effects (ISHE) have stimulated broad interest and debates in the field of spintronics [1-7]. The ability to create a robust pure spin current without magnetic materials is intriguing, and the simplicity and efficiency of this approach is desirable for potential technological applications. Owing to the spin-orbit coupling in heavy nonmagnetic metals such as Pt, a charge current $j$ induces a spin current $j_s$ in the transverse direction. The conversion rate is described by a spin Hall angle $\alpha_H = j_s/j = \sigma_H \rho$, where $\rho$ is the electrical resistivity and $\sigma_H$ is the spin Hall conductivity. For a fixed amount of $j_s$, The Joule heating power density in the material is $j^2 \rho = j_s^2/(\sigma_H^2 \rho)$. Apparently increasing the $\rho$ while maintaining the $\sigma_H$ leads to an enhanced energy efficiency. The spin diffusion length $\lambda$ is also a crucial quantity. For a film that is thicker than $\lambda$, only a depth of $\lambda$ near the surface contribute to the SHE meaningfully.

There are mainly two types of experimental systems to quantify $\alpha_H$ and $\lambda$: the bilayer structure of a ferromagnetic metal and a heavy nonmagnetic metal and the mesoscopic nonlocal structure. The bilayer structures are more actively pursued and allow for a variety of experiments. These include spin pumping [8, 9], spin Hall ferromagnetic resonance [10, 11], spin Hall torque [12-14], and spin Seebeck effect [15]. The less explored mesoscopic nonlocal structure [1-3, 16] takes advantage the nonlocal spin injection and detection methods [17, 18] that involve ferromagnetic electrodes and a nonmagnetic channel (e.g. Cu). Either the SHE or the ISHE can be explored for a heavy nonmagnetic metal (e.g. Pt) that is in contact with the Cu channel.

The bilayer and nonlocal structures involve different physical processes and provide complementary aspects of the SHE/ISHE. In the nonlocal method, the ferromagnetic metal and the Pt are physically separated and therefore other phenomena such as proximity effects or Rashba effect are avoided. The challenge to the nonlocal method, however, is the proper evaluation of various charge and spin transport parameters in the structure. These parameters include spin diffusion length and resistivity values of the Pt film and the Cu channel, the resistance of the Pt/Cu interface, and the spin injection or detection polarization of the ferromagnet. Ex-situ measurements conducted on other samples may not reflect the in-situ values for the SHE/ISHE structures under investigation. In addition, proper quantification of the charge current shunting near the Pt/Cu interface is also crucial and controversy arises from a previous method [19-21].

In this work we explore SHE/ISHE in mesoscopic Pt films using the nonlocal method. The lateral confinement of the films to ~ 230 nm widths gives rise to high electrical resistivity (150 – 300 $\mu\Omega \cdot cm$). The nonlocal SHE/ISHE structures consist of mesoscopic Pt films, Cu channels, Py (permalloy or $Ni_{81}Fe_{19}$ alloy) spin injector/detectors, and low-resistance $AlO_x$ barriers. All relevant physical quantities are either measured directly in the SHE/ISHE structures or determined from in-situ supplementary structures fabricated on the same substrate. Extensive measurements (58 SHE/ISHE structures from 6 substrates with 4 different Pt thickness values) are conducted to take into account of microstructure variations. The charge current shunting and the spin absorption near the $Cu/AlO_x/Pt$ interfaces are consistently characterized by the interfacial resistance. A full quantitative model, based on diffusion equations and proper boundary conditions, is developed to take into account spin transport processes in various materials and interfaces throughout the structure. The effectiveness of the model is demonstrated by extracting the same values of $\alpha_H \lambda$ from two groups of SHE/ISHE structures that differ drastically in the physical sizes and the resistance of the Cu/AlO$_x$/Pt interfaces.

We use the product $\alpha_H \lambda$ as a figure of merit because it is less prone to errors than the individual values of $\alpha_H$ and $\lambda$. Either a larger $\alpha_H$ or a larger $\lambda$ will enhance the transverse spin accumulation on the surface of Pt film that is thicker than $\lambda$. An underestimated $\lambda$ results in overestimated $\alpha_H$, and conversely an overestimated $\lambda$ results in underestimated $\alpha_H$. Furthermore, the $\alpha_H \lambda$ is equivalent to $\sigma_H \rho \lambda$, which is of certain universal quality. Because the $\sigma_H$ is a constant value if the SHE is intrinsic, and the $\rho \lambda$ is also a constant value if the spin relaxation can be described by Elliot-Yafet model with a constant spin-flip probability [22]. Our data and analysis provide an upper limit (0.8 nm) of the $\lambda_{pt}$ and a confident determination of $\alpha_H \lambda_{pt}$ for the resistive mesoscopic Pt films. The substantial value of $\alpha_H \lambda_{pt}$ suggests an efficient spin Hall process.

**II, Sample preparation and measurement**

For each sample, up to 196 mesoscopic metallic structures are fabricated on a 10 mm × 10 mm silicon substrate covered with 200 nm Si$_3$N$_4$. Four different types of structures are included and shown in the scanning electron microscope (SEM) pictures in Fig. 1. Fig 1 (a) and (b) are both nonlocal SHE/ISHE structures. The ~ 280 nm wide ferromagnetic electrode is made of Py and the ~ 85 nm wide nonmagnetic channel is made of Cu. The Pt stripe near the lower end of the Cu channel is ~ 230 nm in width. A directly evaporated AlO$_x$ layer (3nm) is placed at both interfaces forming Py/AlO$_x$/Cu and Cu/AlO$_x$/Pt junctions.

The sizes of the Cu/AlO$_x$/Pt junctions are different for Fig. 1 (a) and (b). Fig 1 (a) illustrates a "small-overlap" SHE/ISHE structure, where the Cu channel and Pt stripe overlaps only near the edge the Pt stripe. The size of the interface is $< 50 \times 30$ nm$^2$, forming a "point-contact". Fig. 1 (b) illustrates a "full-overlap" SHE/ISHE structure, where the overlap size along $x$ direction is comparable to the Pt width forming an interface of area $\sim 80 \times 200$ nm$^2$. The interfacial resistance ($R_i$) values of the small-overlap and full-overlap Cu/AlO$_x$/Pt junctions are quite different, and will be useful in confirming the validity of our quantitative models. The center-to-center distance between the Py/AlO$_x$/Cu and the Pt/AlO$_x$/Cu junctions is defined as channel length $L$.

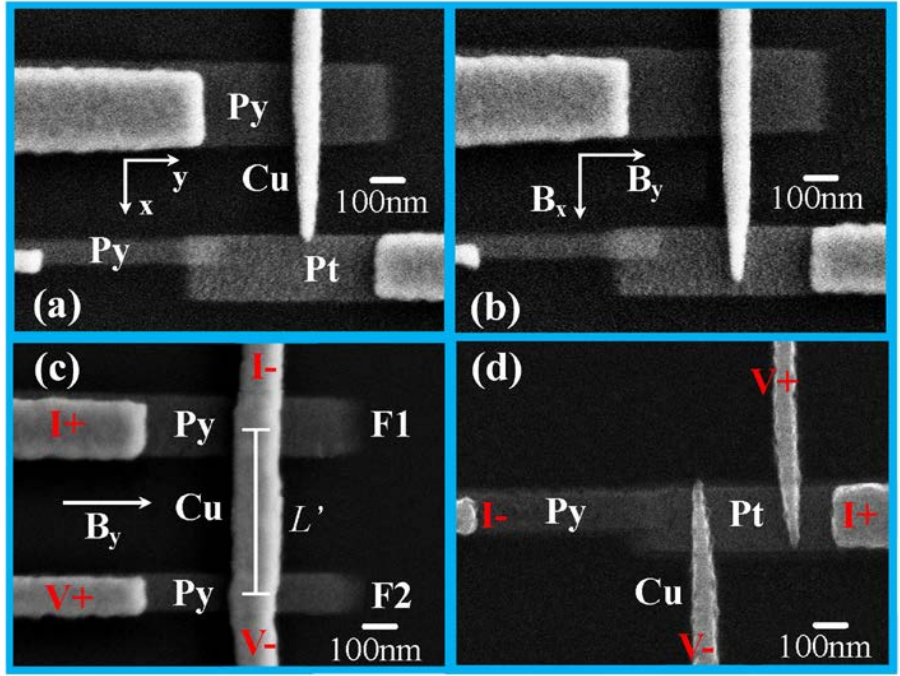

**Figure 1**. SEM pictures of SHE/ISHE structures with (a) small-overlap and (b) full-overlap. (c) SEM image of a nonlocal spin valve and measurement configuration is indicated. $L'$ is the injector-to-detector distance. (d) A Pt resistivity measurement structure. All structures are fabricated on the same substrate through identical processes.

The thickness of Cu and Py is 110 nm and 12 nm, respectively. The relatively large Cu thickness is chosen to ensure long Cu spin diffusion lengths. In this work, Pt thickness of 4 nm, 6 nm, 10 nm and 12 nm has been used for 6 samples (substrates) including a total number of 58 SHE/ISHE structures. We will focus on the results from one sample with 6 nm Pt to illustrate the measurement and quantification method before addressing the dependence on the Pt thickness.

Fig. 1 (c) and (d) illustrates two types of supplementary structures fabricated in-situ on the same substrate with SHE/ISHE structures: the nonlocal spin valve (NLSV) [17, 18, 23-26] and the Pt resistivity structure, respectively. The NLSV structure consists of two Py electrodes (spin injector $F_1$ and spin detector $F_2$) and a Cu channel. AlO$_x$ barriers are placed at the interfaces forming Py/AlO$_x$/Cu junctions. The Pt resistivity structure is a mesoscopic Pt stripe with four electrical probes, with which the resistivity of Pt can be determined. The thickness values of Cu, Py, AlO$_x$, and Pt are the same as those of the SHE/ISHE structures on the same substrate, owing to the identical processes.

For each sample, all structures on the substrate are formed simultaneously by depositing Pt, Py, AlO$_x$, and Cu from different angles through a set of mesoscopic suspended shadow masks, which are created by electron beam (e-beam) lithography from two layers of e-beam resists: the PMGI (polydimethylglutarimide) resist on the bottom and the PMMA (polymethyl methacrylate) resist on the top. Details of the shadow mask and angle deposition method can be found elsewhere [3, 27-29].

The measurements of the 6 samples have been carried out either in a variable temperature probe station or in a pulse-tube variable temperature cryostat. All measurements are conducted at 10 K. The SHE and ISHE measurements from a structure with 6 nm Pt are shown in Fig 2 (a) and (b), respectively, and the corresponding measurement configurations are shown in the insets. For SHE, the current is injected through the Pt stripe (+I on the right and –I on the left), and the nonlocal voltage is detected between Py (+V) and the upper end of Cu channel (-V). For the ISHE, the current flows from Py (+I) to the upper end of Cu (-I), and the nonlocal voltage is measured between the two ends of Pt stripe (+V on the left and –V on the right).

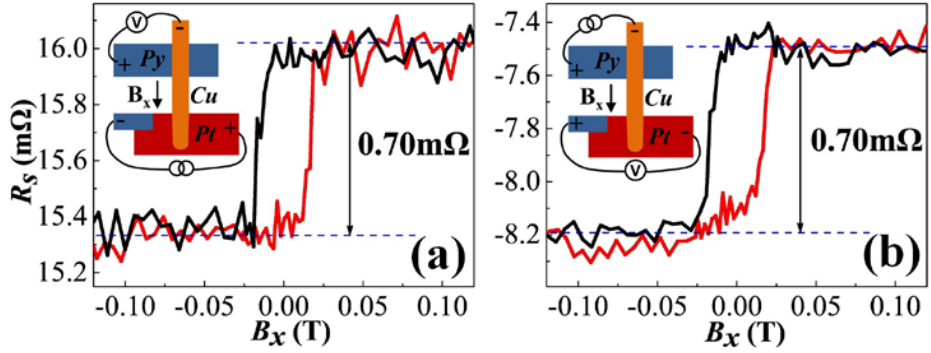

**Figure 2**. The $R_s$ versus $B_x$ curve of (a) a SHE measurement and (b) an ISHE measurement on the same structure with 6 nm Pt. The magnetic field is aligned parallel to Cu channel ($\pm x$ direction) and the temperature is 10 K.

An alternating current (*a.c.*) of $I_e = 0.1$ mA with a frequency of 346.5 Hz is used as the injection current and the nonlocal *a.c.* voltage $V_{nl}$ is detected by a lock-in detector. The nonlocal resistance, $R_s = V_{nl}/I_e$, is recorded as a function of the magnetic field $B_x$ applied along the *x* direction, which is perpendicular to the Py electrodes. In both SHE and ISHE measurements, the $R_s$ value reaches a high state for positive field but reaches a low state for negative field, and the difference between two states is $\Delta R_s = 0.7$ m$\Omega$. The polarity of the signals is consistent with previous SHE/ISHE measurements in Pt. [2, 3] The equal magnitudes of $\Delta R_s$ for SHE and ISHE are expected because of Onsager reciprocal relations. Owing to its better signal-to-noise ratio, the ISHE measurements are used to extract the $\Delta R_s$ values. According to previously used conventions, the SHE/ISHE signal is defined as $\Delta R_{SHE} = \Delta R_s/2$ [2].

The resistance $R_i$ of the Cu/AlOx/Pt interface is individually measured from each SHE/ISHE structure by sending a current from the top end of Cu to the right end of the Pt stripe and detecting a voltage between the Py and the left end of the Pt stripe. Various physical dimensions, as illustrated in Fig. 3 (a) for "small-overlap" and in (b) for "full-overlap", are individually characterized by SEM after the SHE/ISHE measurements. These quantities include width ($w_{pt}$) of Pt stripe, transport distance $L$ between the center of the Py/AlOx/Cu junction and the center of the Cu/AlOx/Pt junction, the width $w_{cu}$ of the Cu channel, the length $d$ of the Cu/AlOx/Pt junction in the $x$ direction, and the average width $w_I$ of the Cu/AlOx/Pt junction in the $y$ direction. Due to the tapering lower end of the Cu channel, typically $w_I < w_{cu}$.

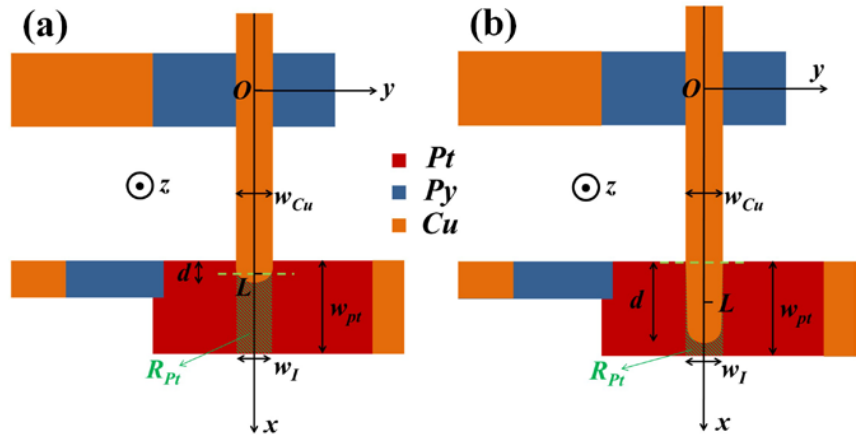

**Figure 3.** Top view of (a) the small-overlap and (b) full-overlap SHE/ISHE structure. The relevant physical dimensions are illustrated.

The effective spin polarization $P$ of the Py/AlO$_x$/Cu interface, the spin diffusion length $\lambda_{cu}$ of the Cu channel, and the resistivity $\rho_{cu}$ of Cu are useful values for quantifying SHE/ISHE structures, and can be derived from the supplementary NLSVs on the same substrate. The $R_s$ versus $B_y$ curve of a NLSV is shown in Fig. 4, and the standard NLSV measurement configuration is illustrated in Fig. 1 (c). The magnetic field $B_y$ is applied along the $y$ direction, which is parallel to the F$_1$ and F$_2$ Py stripes. The field sweep alters the magnetizations of the F$_1$ and F$_2$ between parallel states (high $R_s$) and antiparallel states (low $R_s$), and the difference $\Delta R_s$ is the NLSV spin signal.

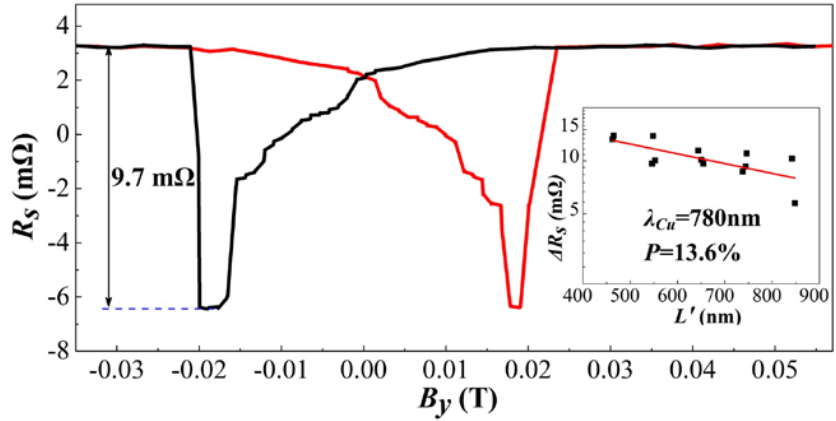

**Figure 4**. The $R_s$ versus $B_y$ curve at 10 K for a NLSV with magnetic field applied parallel to Py electrode ($\pm y$ direction). The $\Delta R_s$ versus $L'$ and a fit (red solid line) is shown in the inset for a sample with 6 nm Pt.

In the inset of Fig. 4, the spin signals $\Delta R_s$ are plotted as a function of the center-to-center distance $L'$ between the F$_1$ and the F$_2$. We fit the $\Delta R_s$ versus $L'$ using $\Delta R_s = (P^2 \rho_{cu} \lambda_{cu}/A_{cu}')exp(-L'/\lambda_{cu})$ to extract the values of $\lambda_{cu}$ and $P$, assuming that the two Py/AlO$_x$/Cu interfaces for F$_1$ and F$_2$ provide the same polarization $P$ [17, 23, 30]. The cross sectional area of the NLSV Cu channel is $A_{cu}' = t_{cu}w_{cu}'$, where $t_{cu} = 110$ nm and $w_{cu}'$ are the Cu thickness and width, respectively, and $w_{cu}'$ is measured by SEM. Here for the NLSVs, $L'$ and $w_{cu}'$ are used to be distinguished from the channel length $L$ and Cu width $w_{cu}$ of the SHE/ISHE structures. In this set of data the fitting yields $P = (13.6 \pm 2.6)$ % and $\lambda_{cu} = (780 \pm 220)$ nm at 10 K.

As described in other works, the precise determination of either $P$ or $\lambda_{cu}$ is nontrivial [31, 32]. The overestimate of one value leads to the underestimate of the other, and vice versa. However, the truly relevant quantity for later analysis (in Eq. 3 and 4) is the spin current that flows down the channel and it scales with $P\exp(-L/\lambda_{cu})$. This quantity is less uncertain than the individual values of $P$ or $\lambda_{cu}$ [31]. From the data in the inset of Fig. 4, this quantity for $L = 500$ nm is determined to be $P\exp(-L/\lambda_{cu}) = (0.072 \pm 0.005)$, which has less uncertainty than individual values of $P$ or $\lambda_{cu}$.

To obtain the resistivity value of Cu, the Cu resistance $R_{cu}$ can be determined by sending in a current through the Cu channel and detecting voltages between F$_1$ and F$_2$. Then the resistivity $\rho_{cu}$ can be calculated from $R_{cu}$, $L'$ and $A_{cu}'$. For each sample, 10 - 15 NLSVs are used to obtain the $P$, $\lambda_{cu}$, and $\rho_{cu}$.

The resistivity of mesoscopic Pt stripes is measured using the supplementary structures shown in Fig. 1 (d). Note that the widths of these Pt stripes are the same as those in the SHE/ISHE structures. For each sample, 5 – 10 Pt stripes are measured for resistivity and the average value is used as the $\rho_{pt}$ of that sample. The values of $\rho_{pt}$ are generally in the range of 150 μΩ•cm < $\rho_{pt}$ < 300 μΩ•cm, and this is a factor of 5 to 10 larger than that of extended films. The large resistivity is due to the reduced lateral dimension and thickness. The $\Delta R_{SHE}$, $R_i$, $L$, $w_{pt}$, $w_{cu}$, and $w_I$ values from each SHE/ISHE structure and the values of $P$, $\lambda_{cu}$, $\rho_{cu}$, and $\rho_{pt}$ from supplementary structures will be used for quantitative analysis of ISHE signals.

Fig. 5 (a) and (b) shows representative ISHE measurements for a small-overlap structure ($\Delta R_s$ = 0.41 m$\Omega$) and a full-overlap structure ($\Delta R_s$ = 0.83 m$\Omega$), respectively, with magnetic field along the *x* direction. The ISHE measurements with a magnetic field along *y* direction (parallel to Py injectors) are shown in Fig. 5 (c) and (d) for the same two structures as in Fig. 5 (a) and (b), respectively. At large ± *B*, spin moments along ± *y* are injected and the ISHE voltage generated in the Pt film is in the *x* direction. Therefore the *R*s value measured across the Pt stripe, which lies in the *y* direction, reaches the same value for large + *B* and - *B*. At intermediate fields, the magnetization of Py rotates in the substrate plane and has a non-zero projection on the *x* axis. Therefore a variation of *R*s is observed across the Pt stripe. However the $\Delta R_s$ values are smaller than those in Fig. 5 (a) and (b), because the Py magnetization is never fully aligned along ± *x* direction. This is consistent with our previously published results [3].

The $\Delta R_s$ values of all SHE/ISHE structures on this substrate are plotted as a function of the channel length *L* in Fig. 5 (e). The average $\Delta R_s$ of the full-overlap structures is 0.63 m$\Omega$, which is 1.8 times of the average $\Delta R_s$ of 0.35 m$\Omega$ for the small-overlap structures. The solid line is a fit assuming an exponential decay of the $\Delta R_s$ as a function of *L* with $\lambda_{cu}$ as the decay length. The resistance values *R*i of the Cu/AlO$_x$/Pt interfaces are shown for SHE/ISHE structures with various channel length *L* in Fig. 5 (f). The *L* values are used as a labeling mechanism for various SHE/ISHE structures without implying any necessary correlation between *R*i and *L*. The small-overlap structures have a larger average *R*i of 85 $\Omega$ than that (3 $\Omega$) of the full-overlap structures.

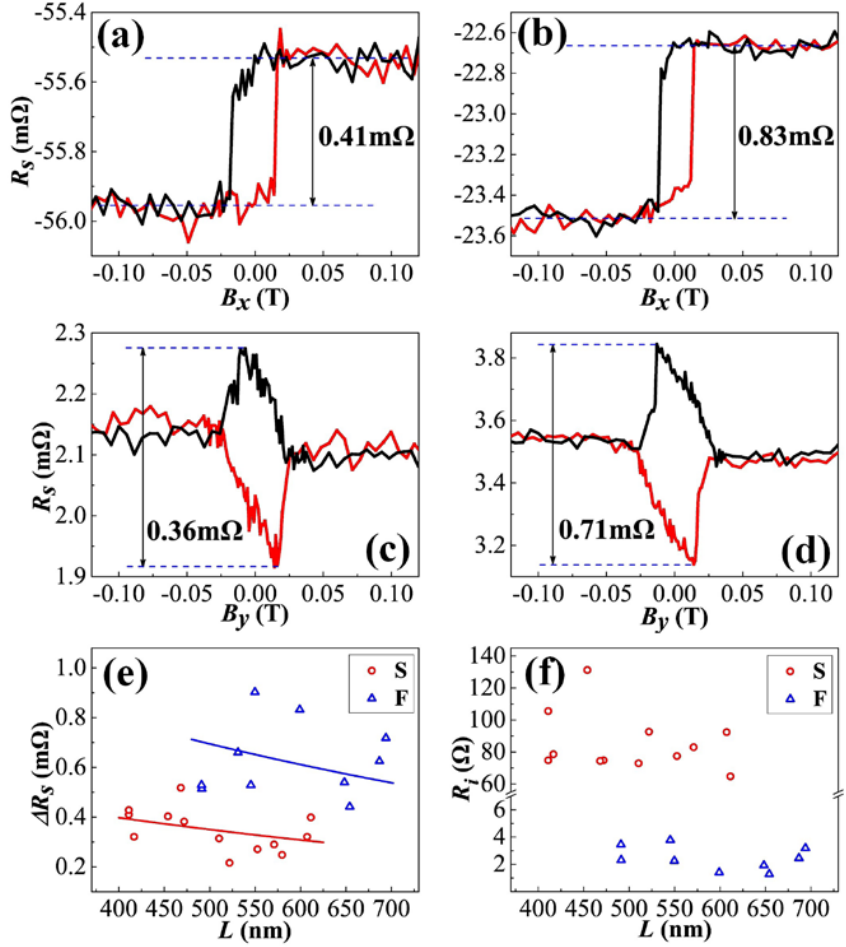

**Figure 5**. Various measurements at 10 K from a sample with 6 nm Pt. ISHE measurements on (a) a small-overlap structure and (b) a full-overlap structure with field along *x* direction. ISHE measurements on (c) a small-overlap structure and (d) a full-overlap structure with field along *y* direction. (e) The $\Delta R_s$ values from ISHE measurement of field along $\pm x$ direction versus channel length *L* of the SHE/ISHE structures. (f) The resistance of the Cu/AlO$_x$/Pt interface for SHE/ISHE devices identified by *L*.

**III, Quantitative analysis of SHE/ISHE signals**

In this section, we use simple models to derive a relationship between the magnitude of $\Delta R_s = 2\Delta R_{SHE}$, the spin Hall angle $\alpha_H$, the spin diffusion length $\lambda_{pt}$ of Pt, and other measureable quantities in SHE/ISHE structures. This will allow us to obtain the $\alpha_H$ and $\lambda_{pt}$ from the experimental results. The calculation has been carried out with both SHE and ISHE and the results are consistent. The models yield the same average spin Hall angles for the small-overlap and full-overlap structures, attesting to the effectiveness of the models.

Refer to Fig. 3 (a) and (b) for relevant dimensions. In the context of ISHE, a spin current is injected from Py and flows down the Cu channel along the $+x$ direction. Upon reaching the Cu/AlO$_x$/Pt interface near the end of the Cu channel, a reflected spin current flows toward $-x$ direction and an absorbed spin current flows across the AlO$_x$ into the Pt. The absorbed spin current flows perpendicularly into the Pt film along $z$ direction and gives rise to the ISHE voltage.

From one-dimensional diffusion equation, the spin accumulation in Cu is described as $\delta\mu_{cu}(x) = a_1 exp(-x/\lambda_{cu}) + a_2 exp(x/\lambda_{cu})$ for $0 < x < L$ and $\delta\mu_{cu}(x) = a_1' exp(x/\lambda_{cu})$ for $x < 0$, where $\delta\mu_{cu} = \mu_{cu\uparrow} - \mu_{cu\downarrow}$. The combined electrochemical potential $\mu_{cu\uparrow,\downarrow}$ is defined as $\mu_{cu\uparrow,\downarrow} = -\mu_{ch\uparrow,\downarrow}/e + V$, where $\mu_{ch\uparrow,\downarrow}$ is the chemical potential for spin-up (-down) and $V$ is the electrical voltage. The spin current in the Cu channel is $I_s(x) = -\frac{\sigma_{cu}}{2}\left(\frac{d\delta\mu_{cu}(x)}{dx}\right)A_{cu}$, where $\sigma_{cu}$ is the Cu conductivity and $A_{cu}$ is the cross sectional area of the Cu channel. At $x = 0$, boundary conditions are $\delta\mu_{cu}(0^-) = \delta\mu_{cu}(0^+)$ and $I_s(0^+) - I_s(0^-) = PI_e$, where $I_e$ is the charge injection current through the Py/AlO$_x$/Cu interface and $P$ is the effective injection polarization of the interface. Note that at $x < 0$, the spin current is toward $-x$ direction and therefore carries a negative value. At $x = L$, boundary conditions are $I_s(L) = I_{sa}$ and $\frac{\delta\mu_{cu}(L)}{2R_i} = I_{sa}$, where $I_{sa}$ is the absorbed spin current into the Pt. The latter equation indicates that the $I_{sa}$ is driven by the spin accumulation difference across the Cu/AlO$_x$/Pt interface, which has a resistance of $R_i$. We obtain the absorbed spin current: $I_{sa} = 1/2 \gamma P I_e exp(-L/\lambda_{cu})$, where

$$\gamma = 2R_{scu}/(R_{scu} + R_i) \qquad (1)$$

is the spin absorption coefficient and $R_{scu} = \lambda_{cu}/\sigma_{cu}A_{cu} = \rho_{cu}\lambda_{cu}/A_{cu}$ is the spin resistance of the Cu channel. If $R_i = 0$, $\gamma = 2$; if $R_i \gg R_{scu}$, $\gamma = 2R_{scu}/R_i \ll 1$.

The absorbed spin current flows perpendicularly (along $z$ direction) into Pt and the ISHE develops a voltage in the $y$ direction. Consider $z = 0$ at the top surface of the Pt film and $z = t_{pt}$ for the bottom surface of the Pt film. The spin current density absorbed into the Pt top surface is $j_{sa} = I_{sa}/A_j$, where $A_j$ is the area of the Cu/AlOx/Pt junction. If assuming uniform junction width $w_I$, $A_j = w_I d$, where $d$ is the overlap between Cu and Pt along the $x$ direction. The spin injection into Pt induces a spin accumulation, which can be described by $\delta\mu_{pt}(z) = u_1 exp(-z/\lambda_{pt}) + u_2 exp(z/\lambda_{pt})$. The spin current density in the z direction is $j_s(z) = -\frac{\sigma_{pt}}{2}\left(\frac{d\delta\mu_{pt}(z)}{dz}\right)$, where $\sigma_{pt} = 1/\rho_{pt}$ is the Pt conductivity. Boundary conditions are $j_s(0) = j_{sa}$ and $j_s(t_{pt}) = 0$, which states that the spin current near the top surface equal to the absorbed spin current and the spin current near the bottom surface vanishes. Then the coefficients $u_1$ and $u_2$ can be solved from the boundary conditions and the $j_s(z)$ is determined. The ISHE induces a charge current $\alpha_H j_s(z)$ in the y direction and therefore an electric field $E_y(z) = \alpha_H \rho_{pt} j_s(z)$. The resulting voltage between two ends of the Pt stripe is $V_{nl} = w_I \overline{E_y(z)} = w_I \frac{1}{t_{pt}} \int_0^{t_{pt}} E_y(z) dz$.

The voltage on the Pt stripe changes sign and becomes $-V_{nl}$ when the injected spins at $x = 0$ from the Py changes sign. Therefore $\Delta R_s = \Delta V_{nl}/I_e = 2V_{nl}/I_e$. In addition, the shunting effect from two sources will reduce the $V_{nl}$. The Pt at $x > L + d/2$ is not in contact with the Cu channel, does not receive a spin current on the top surface, and therefore generates no ISHE voltage. As a result, the $V_{nl}$ should be multiplied by a reduction factor $d/w_{pt}$. Also, the highly conductive Cu channel shunts the ISHE voltage through the low-resistance oxide barrier and the reduction factor is

$$\chi = 4R_i/(4R_i + R_{pt}) \qquad (2),$$

where $R_{pt} = \rho_{pt} w_I/(w_{pt} t_{pt})$ is the resistance along the $y$ direction of the Pt segment that is shown as green shaded area in Fig. 3 (a), (b) and Fig. 6 (a). Parallel to $R_{pt}$ is another conduction channel that passes through half of the Cu/AlOx/Pt junction, the highly conductive Cu, and the other half of the Cu/AlOx/Pt junction, as shown in Fig. 6 (a) and (b). The resistance of this parallel channel is $4R_i$, since each half-junction has the resistance value of $2R_i$ and the Cu resistance (along the $y$ direction) is negligible compared to $R_i$. The Pt segment can be seen as an electromotive force (ISHE voltage) with an internal resistance $R_{pt}$. The resistance of $4R_i$ can be considered as the external resistance, and the actually measured voltage should be the terminal voltage on the $4R_i$. The ratio $\chi$ between the terminal voltage and the emf voltage is therefore expressed by Eq. 2. Summarizing all above, the ISHE signal can be calculated:

$$\Delta R_s = 2\Delta R_{SHE} = \frac{\alpha_H \gamma \chi \lambda_{pt} \rho_{pt} P}{t_{pt} w_{pt}} \left( \frac{exp(t_{pt}/\lambda_{pt})-1}{exp(t_{pt}/\lambda_{pt})+1} \right) exp(-L/\lambda_{cu}) \qquad (3).$$

Note that the shape of the lower tip of the Cu channel does not affect the above result. To prove this, we assume a variable width along $x$ direction $w_i(x)$ instead of a constant $w_I$, and thereby $V_{nl}$ depends on the $x$ as well: $V_{nl}(x) = w_i(x)\overline{E_y(z)} \propto w_i(x)\overline{J_s(z)} \propto w_i(x)\frac{I_{sa}}{A_j}$, where $A_j = \int_{L-d/2}^{L-d/2} w_i(x)dx$ is the area of the Cu/AlOx/Pt
15

junction. The measured voltage should be an average over $x$: $\overline{V_{nl}} \propto \frac{I_{sa}}{A_j}$.

$\frac{1}{d}\int_{L-d/2}^{L+d/2} w_{cu}(x)dx = \frac{I_{sa}}{d}$, which is independent of $A_j$ or a particular form of $w_i(x)$ [3].

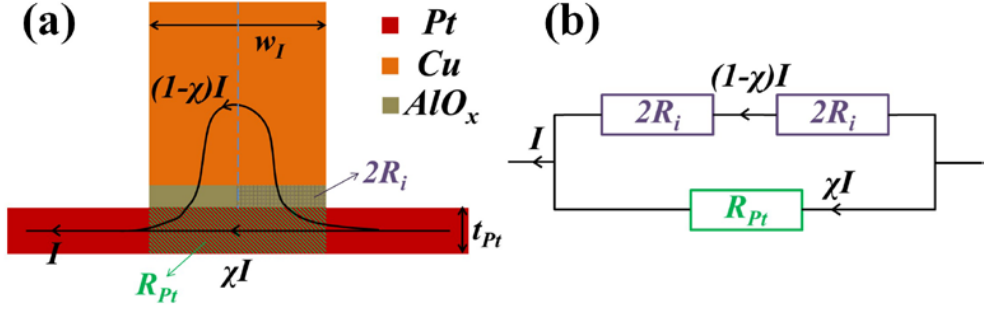

**Figure 6.** (a) The cross-sectional view of Fig. 3 (a) or (b) along the green dashed line perpendicular to the Cu channel and the distribution of charge current for SHE measurement. (b) Resistor model for calculating shunting factor $\chi$.

The calculation has also been done in the context of SHE and gives the exact same expression for $\Delta R_s$ as in Eq. 3. This is expected from the reciprocal relationship between SHE and ISHE and reassures the validity of Eq. 3. In the SHE calculation, spin accumulation in Pt and Cu satisfies the diffusion equations, and the interfacial spin current across the Cu/AlOx/Pt is driven by the difference of the spin accumulation between two sides of the interface. Near the top and bottom Pt surfaces, the spin current driven by the spin accumulation cancels the spin current driven by the SHE in the z direction.

The factor $\chi$ can be understood in a more straightforward manner in the SHE, as shown in Fig. 6 (a), which is a cross-sectional view of the structures in Fig. 3(a) or (b) along the dashed line. When a charge current is sent through the two ends of the Pt stripe, only a fraction ($\chi$) of the current stays in the Pt film, and the rest ($1 - \chi$) of the current is shunted by the Cu. As in Fig. 6 (a), the shunted current flows across right-half of the AlO$_x$ interface (with resistance $2R_i$) into the Cu and then flows across the left-half of the AlO$_x$ interface (with resistance $2R_i$) out of Cu. Therefore the resistance to the shunted current is $4R_i$, neglecting the resistance of the highly conductive Cu. The resistance to the current in the parallel branch through Pt is $R_{pt}$. The equivalent circuit model is shown in Fig. 6 (b). From current divider rule for parallel resistors, we have obtained the expression of $\chi$ in Eq. 2.

**IV Determination of $\alpha_H \lambda_{pt}$**

The expression of $\Delta R_s$ in Eq. 3 can be rewritten as:

$$\Delta R_s = 2\Delta R_{SHE} = \frac{\alpha'_H \gamma \chi \rho_{pt} P}{2w_{pt}} exp\left(-\frac{L}{\lambda_{cu}}\right) \qquad (4)$$

with the definition of an apparent spin Hall angle $\alpha'_H$:

$$\alpha'_H(t_{pt}) = 2\alpha_H \left(\frac{\lambda_{pt}}{t_{pt}}\right)\left(\frac{exp(t_{pt}/\lambda_{pt})-1}{exp(t_{pt}/\lambda_{pt})+1}\right) \qquad (5).$$

Note that $\alpha'_H$ is a mere definition for convenience and its value monotonically decays as a function of the Pt thickness $t_{pt}$. But $\alpha_H$ is the real Pt spin Hall angle and does not depend on the $t_{pt}$. In the limit of thin Pt films (*i.e.* $t_{pt} \ll \lambda_{pt}$), the apparent spin Hall angle is a constant and shares the same value with the spin Hall angle: $\alpha'_H = \alpha_H$. Therefore, in this limit, the Pt spin Hall angle $\alpha_H$ can be directly calculated from the measured signal $\Delta R_s$ by using Eq. 4. As $t_{pt}$ increases above the thin limit, the $\alpha'_H$ monotonically but gradually decreases, reaching $\alpha'_H = 0.76\alpha_H$ when $t_{pt} = 2\lambda_{pt}$. In the limit of thick Pt films (*i.e.* $t_{pt} \gg \lambda_{pt}$), $\alpha'_H(t_{pt}) = 2\alpha_H \left(\frac{\lambda_{pt}}{t_{pt}}\right)$ and the apparent spin Hall angle $\alpha'_H$ is inversely proportional to the Pt film thickness. Because only a thickness of $\lambda_{pt}$ near the top Pt surface contributes to the SHE signal. As a result, in this thick limit, the $\alpha'_H$ decays more rapidly as a function of $t_{pt}$. The magnitude of $\Delta R_s$ essentially follows a similar dependence on $t_{pt}$, assuming that other physical parameters are fixed. For an unknown relationship between $\lambda_{pt}$ and $t_{pt}$, one can first use Eq. 4 to calculate the apparent spin Hall angle $\alpha'_H$ from experimental values of $\Delta R_s$ for samples with various $t_{pt}$ values. Then a fit of the $\alpha'_H$ versus $t_{pt}$ dependence by Eq. 5 will generate the Pt spin Hall angle $\alpha_H$ and the Pt spin diffusion length $\lambda_{pt}$.

We apply this method to our experimental data. Table 1 lists values of $\Delta R_s$, $R_i$, $\gamma$, $\chi$, and $\alpha'_H$ for several SHE/ISHE structures with 6nm Pt film. The three small overlap structures (S-4, S-6, and S-12) has larger $\chi$ but smaller $\gamma$ compared to the full overlap structures (F-1, F-4, and F-12). The larger junction resistance $R_i$ of the small-overlap reduces the shunting effect and therefore allows a higher fraction ($\chi$) of current to remain in Pt (in the context of SHE). The spin current across the interface is also reduced owing to the larger $R_i$ and therefore $\gamma$ is reduced. The two important processes, the shunting of charge current and the transport of interfacial spin current, are both effectively quantified by the junction resistance $R_i$ through Eq. 1 and 2.

| Device | $\Delta R_s$ | $L$ | $R_i$ | $\chi$ | $\gamma$ | $\alpha_H$' |
|---|---|---|---|---|---|---|
| | m$\Omega$ | nm | $\Omega$ | | | |
| S-4 | 0.38 | 471.9 | 74.8 | 0.80 | 0.033 | 0.040 |
| S-6 | 0.52 | 468.2 | 74.5 | 0.79 | 0.033 | 0.052 |
| S-12 | 0.27 | 552.6 | 77.5 | 0.82 | 0.032 | 0.032 |
| F-1 | 0.51 | 491.3 | 2.3 | 0.077 | 0.64 | 0.030 |
| F-4 | 0.90 | 549.7 | 2.2 | 0.074 | 0.66 | 0.058 |
| F-12 | 0.54 | 648.3 | 1.9 | 0.066 | 0.71 | 0.040 |

**Table 1.** The $\Delta R_s$ of ISHE, channel distance $L$, interface resistance $R_i$ and calculated shunting factor $\chi$, spin absorption rate $\gamma$, and apparent spin Hall angle $\alpha_H$' for selected small-overlap (S-4, 6, 12) and full-overlap SHE/ISHE structures (F-1 , 4, 12) for a sample with $t_{pt}$ = 6 nm.

The values of the apparent spin Hall angle $\alpha'_H$ calculated from Eq. 4 from small-overlap structures with various $L$ values are plotted in Fig. 7 (a) and those from the full-overlap are plotted in Fig. 7 (b). All values come from SHE/ISHE structures with 6 nm Pt on the same substrate. Interestingly, the average $\alpha'_H$ values are exactly the same: $\alpha_H$' = (0.043 ± 0.011) for small-overlap and $\alpha_H$' = (0.043 ± 0.013) for full-overlap, despite the difference of average $R_i$ by a factor of 28 and the difference of average $\Delta R_s$ by a factor of 1.8. This attests to the validity and consistency of this method.

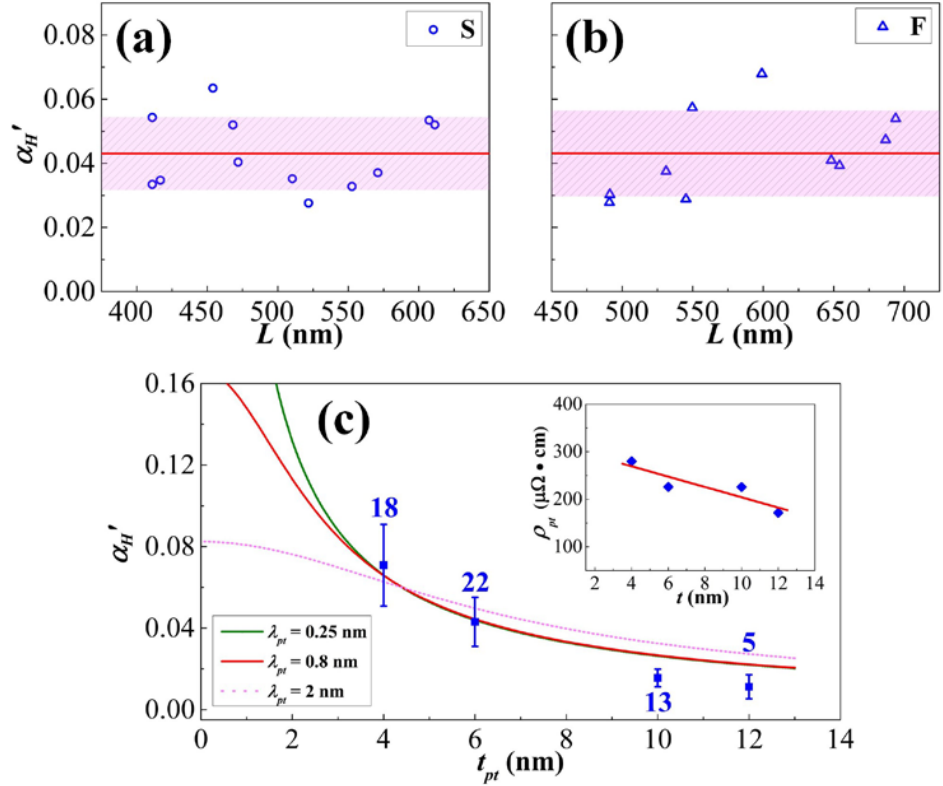

**Figure 7.** Calculated apparent spin Hall angle $\alpha_H$' for (a) the small-overlap structures and (b) the full-overlap structures from the sample with 6 nm Pt at 10K. The red lines indicate average values in each case and the shaded areas indicate the standard deviation. (c) The value of $\alpha_H$' as a function of Pt thickness from 58 SHE/ISHE devices. The number next to each data point indicates the number of SHE/ISHE structures measured for that thickness. The solid and dashed lines are fits with various $\lambda_{pt}$. The fitting is weighted by the number of structures at each thickness. The inset of (c) shows the Pt resistivity as a function of Pt thickness.

Using this method, the $\alpha_H$' values from samples with different Pt thickness are determined and plotted as a function of the Pt thickness, as shown in Fig. 7 (c). A total number of 58 SHE/ISHE structures with 4 different $t_{pt}$ values from 6 substrates are included. As $t_{pt}$ increases from 4 nm to 12 nm, a drastic decay of $\alpha_H'$ is observed and indicates that $t_{pt} > \lambda_{pt}$. Because when $t_{pt} < \lambda_{pt}$, $\alpha_H'$ should remain nearly a constant and be approximately equal to $\alpha_H$ according to Eq. 5. When a fit of the $\alpha_H'$ versus $t_{pt}$ by Eq. 5 is conducted with $\alpha_H$ and $\lambda_{pt}$ as free parameters, the best fit generates short spin diffusion length $\lambda_{pt} = 0.25$ nm and large spin Hall angle $\alpha_H = 0.53$. The fitted curve is shown as the green line in Fig 7 (c). The product of the two values is $\alpha_H \lambda_{pt} = 0.132$ nm. Note that the fitting is weighted by the number of SHE/ISHE structures investigated at each thickness. This number is indicated next to each data point in Fig. 7 (c).

We also used other fixed values of $\lambda_{pt}$ within the range of 0.25 nm $\leq \lambda_{pt} \leq$ 0.8 nm and left $\alpha_H$ as the single fitting parameter. Interestingly we could obtain equally satisfactory fits for any $\lambda_{pt}$ in that range. Though the fitted $\alpha_H$ decreases as the assumed $\lambda_{pt}$ increases, the product of the two always remains the same. The red line in Fig. 7 (c) corresponds to $\lambda_{pt} = 0.8$ nm and $\alpha_H = 0.167$, yielding $\alpha_H \lambda_{pt} = 0.133$ nm. Within the range of the experimental data (4 nm ≤ $t_{pt}$ ≤ 12 nm), this curve is almost identical as the green curve ($\lambda_{pt}$ = 0.25 nm), and both scale with $1/t_{pt}$. The difference between two curves lies in the region $t_{pt}$ < 4.0 nm, where no experimental data is present.

The experimental $\alpha'_H$ values for 10 nm and 12 nm are below the fitted curves, because the decay of $\alpha'_H$ as a function of $t_{pt}$ is faster than the $1/t_{pt}$ trend given by the model. For $\lambda_{pt}$ > 0.8 nm, the calculated curves show even slower decaying trend, and obviously cannot describe the experimental data well. Fig. 7 (c) shows a fitting curve for $\lambda_{pt}$ = 2.0 nm and its correlation with experimental data is clearly worse than the curve with $\lambda_{pt}$ < 0.8 nm. Therefore, we can safely conclude that the upper limit of the $\lambda_{pt}$ is 0.8 nm. The present set of experimental data cannot conclude the precise value of $\lambda_{pt}$, but the fitting with various $\lambda_{pt}$ values between 0.25 nm and 0.8 nm gives a consistent product of $\alpha_H \lambda_{pt}$ = (0.133 ± 0.067) nm.

The short $\lambda_{pt}$ is consistent with the high resistivity of the mesoscopic Pt film and with the works by Liu *et al.* [13], Zhang *et al.* [33], and Nguyen *et al* [22]. The $\rho_{pt}$ as a function of film thickness is shown in the inset of Fig. 7 (c), and shows gradual decrease as the *t*$_{pt}$ is increased. From Eq. 4 and 5, it is obvious that the SHE/ISHE signal Δ*R*$_{SHE}$ is proportional to $\alpha_H \rho_{pt}$ or equivalently $\sigma_H \rho_{pt}^2$. An underestimated *ρ*$_{pt}$ would lead to an overestimated spin Hall angle $\alpha_H$ or spin Hall conductivity $\sigma_H$. Therefore the accurate determination of the in-situ resistivity is an essential component of quantifying the SHE/ISHE. Using the average $\rho_{pt}$ = 225 μΩ•cm from the inset of Fig. 7 (c) and 0.25 nm < $\lambda_{pt}$ < 0.8 nm, we have 0.56 × 10$^{-15}$ Ω•m² < $\rho_{pt} \lambda_{pt}$ < 1.8 × 10$^{-15}$ Ω•m². Using $\alpha_H \lambda_{pt}$ = $\sigma_H \rho_{pt} \lambda_{pt}$ = 0.133 nm, we estimate the spin Hall conductivity to be 0.74 ×10$^5$ Ω$^{-1}$m$^{-1}$< $\sigma_H$

< $2.4 \times 10^5$ $\Omega^{-1}$m$^{-1}$. As a comparison, Nguyen et al. [22] obtained $\rho_{pt}\lambda_{pt} = (0.77 \pm 0.08) \times 10^{-15}$ $\Omega \cdot$m$^2$ and $\sigma_H = (5.9 \pm 0.2) \times 10^5$ $\Omega^{-1}$m$^{-1}$ from spin Hall torque measurements.

The electron mean free path in the Pt films can be estimated to be 0.21 nm from the average $\rho_{pt}$ using the Drude model. This value is lower than the range of spin diffusion length (0.25 nm < $\lambda_{pt}$ < 0.8 nm), as expected from the Elliott-Yafet spin relaxation mechanism. However, the short $\lambda_{pt}$ (< 0.8 nm) suggests that the effect is more sensitive to the surface region of the film. As a point of reference, the lattice constant of Pt is 0.39 nm.

Since we are confident that the experimental results are in the limit of $t_{pt} \gg \lambda_{pt}$, we obtain $\alpha_H \lambda_{pt} = \alpha'_H(t_{pt}/2)$ from the limiting form of Eq. 5. Therefore, the $\alpha_H \lambda_{pt}$ can be calculated for each SHE/ISHE structure from the $\alpha'_H$. The results obtained from all 58 devices are summarized in Fig. 8 (a), where the value is plotted against the $L$ value of each structure. Again, $L$ is used as a labeling mechanism without suggesting any necessary dependence of $\alpha_H \lambda_{pt}$ on $L$. Different Pt thicknesses $t_{pt}$ are represented by different symbols. Though the average of all 58 structures is $\alpha_H \lambda_{pt} = (0.12 \pm 0.05)$ nm, the data are scattered over a broad range. A substantial number (15 out of 58) of structures show 0.15 nm < $\alpha_H \lambda_{pt}$ < 0.21 nm. The highest value is $\alpha_H \lambda_{pt} = 0.21$ nm for a structure with 4 nm Pt.

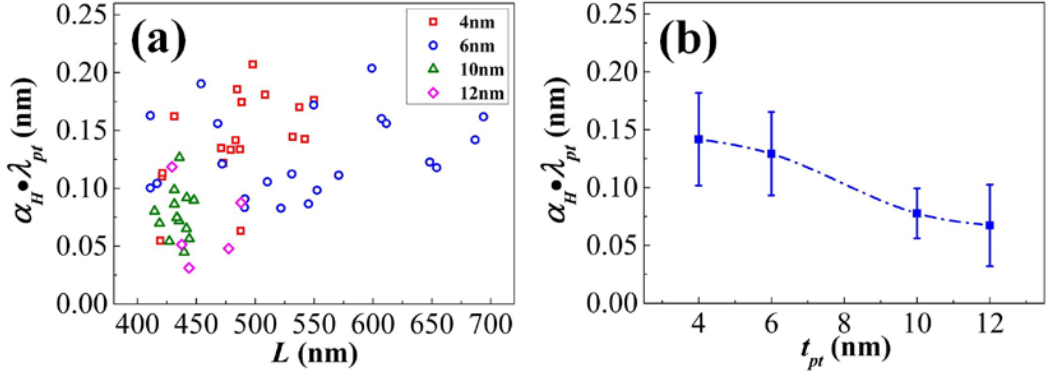

**Figure 8**. (a) Calculated $\alpha_H \lambda_{pt}$ for each SHE/ISHE structure is plotted against the channel distance $L$ of the structure. Different symbols represent various thicknesses. (b) The average experimental value of $\alpha_H \lambda_{pt}$ is plotted as a function of $t_{pt}$. The dash-dot line is a guidance of eyes.

It is noticeable in Fig. 8 (a) that the 4 nm and 6 nm Pt films tend to show higher $\alpha_H \lambda_{pt}$ than the 10 nm and 12 nm Pt films. Therefore we plot the average $\alpha_H \lambda_{pt}$ of each Pt thickness as a function of $t_{pt}$ in Fig. 8 (b). We have obtained $\alpha_H \lambda_{pt} = (0.142 \pm 0.040)$ nm for 4 nm Pt, $\alpha_H \lambda_{pt} = (0.129 \pm 0.036)$ for 6 nm Pt, $\alpha_H \lambda_{pt} = (0.078 \pm 0.022)$ for 10 nm Pt, and $\alpha_H \lambda_{pt} = (0.067 \pm 0.035)$ for 12 nm Pt. There is a gradually decreasing trend of $\alpha_H \lambda_{pt}$ as Pt thickness $t_{pt}$ increases. The dash-dot line is a guidance of eyes. As stated earlier, the quantity $\alpha_H \lambda_{pt}$ is equivalent to $\sigma_H \rho_{pt} \lambda_{pt}$, which is supposed to be a constant if the spin Hall effect is intrinsic (constant $\sigma_H$) and if the spin relaxation in Pt can be described by Elliott-Yafet model (constant $\rho_{pt} \lambda_{pt}$). Therefore the decreasing trend in Fig. 8 (b) suggests that either the SHE is not entirely intrinsic or the assumption of constant $\rho_{pt} \lambda_{pt}$ is oversimplified. It is unclear at this point which is the primary cause.

**V, Further discussions**

The results (values of $\sigma_H$ and $\rho_{pt}\lambda_{pt}$) by Nguyen *et al*. [22] lead to a large value of $\sigma_H\rho_{pt}\lambda_{pt} = 0.45$ nm. Our result of average $\sigma_H\rho_{pt}\lambda_{pt} = (0.142 \pm 0.040)$ nm for 4 nm Pt is lower by more than a factor of 3. However, our highest $\sigma_H\rho_{pt}\lambda_{pt}$ value from an individual SHE/ISHE structure is 0.21 nm, roughly half of the value by Nguyen *et al*. By using spin pumping method, Feng *et al*. [34] reported $\alpha_H = 0.012$ and $\lambda_{pt} = 8.3$ nm, yielding $\alpha_H\lambda_{pt} = 0.10$ nm. The same $\alpha_H\lambda_{pt} = 0.10$ nm can be inferred from the spin pumping measurements by Zhang *et al.* [33], but the values of $\alpha_H = 0.086$ and $\lambda_{pt} = 1.2$ nm are different. Qu *et al*. [35] used spin Seebeck effects and ISHE, and reported $\alpha_H = 0.013$ and $\lambda_{pt} = 2.5$ nm and thereby $\alpha_H\lambda_{pt} = 0.03$ nm.

A comparison can also be made with nonlocal measurements on mesoscopic Pt films by other groups. Morota *et al*. [20] reported $\lambda_{pt} = 11$ nm and $\alpha_H = 0.021$, yielding $\alpha_H\lambda_{pt} = 0.23$ nm. Though this value is in reasonable agreement with ours, the $\lambda_{pt}$ is much higher and the $\alpha_H$ is much lower than our estimation. In addition, it is noteworthy to point out that the reported resistivity for their mesoscopic Pt films is $\rho_{pt} = 12.3$ μΩ•cm, which is much lower than the values measured in-situ from our mesoscopic Pt films. Isasa *et al*. [36] reported $\lambda_{pt} = 3.4$ nm and $\alpha_H = 0.009$, yielding $\alpha_H\lambda_{pt} = 0.03$ nm, which is much smaller than our value. Their reported resistivity value is $\rho_{pt} = 25$ μΩ•cm.

The broad distribution of the results in the literature is not completely surprising. Our own results in Fig. 8 (a) scatter broadly between 0.03 nm and 0.21 nm. The microstructure of the Pt films likely imposes strong influences on the SHE. The salient difference between our experiments and others is that our Pt films are truly mesoscopic with high resistivity and therefore a short $\lambda_{pt}$ can be confidently concluded. The microstructures of the mesoscopic films may vary and induce variations of $\sigma_H$, $\rho_{pt}$, and $\lambda_{pt}$ between individual structures. In contrast, effects measured from larger films may represent average behavior over a large area. Overall, the $\sigma_H \rho_{pt} \lambda_{pt}$ values we obtained are still quite substantial in magnitudes.

Notes should be given to our previous work in Ref. 3. Comparing Eq. 4 in this paper to Eq. 1 of Ref. 3, there is a difference by a factor 2. In the previous work, we assumed that the absorbed spin current into Pt is uniform in the z direction. However, as discussed in the previous section, the spin current $j_s(z)$ in Pt is not uniform and reaches zero at the bottom surface of the Pt film ($z = t_{pt}$). In Ref. 3, we reported a resistivity of 26 μΩ•cm, which was measured from an extended Pt film of 6 nm thickness. The actual in-situ resistivity from the mesoscopic Pt film of 6 nm thickness should be much larger and be comparable to the values reported in this work.

## VI, Conclusions

In conclusion, we use nonlocal structures to demonstrate enhanced spin Hall effects and inverse spin Hall effects in mesoscopic Pt films. Essential physical quantities are all determined in-situ on the same substrate and provide an accurate representation of the structures. The resistivity of the mesoscopic Pt films is substantially higher than extended Pt films and can be beneficial for the energy efficiency of the spin Hall effects. The spin absorption into Pt and the current shunting by Cu are treated effectively using simple models with resistors and spin resistors. By consistent analysis of the samples with various Pt thicknesses, we confidently set an upper limit of 0.8 nm for the Pt spin diffusion length.

The product of the spin Hall angle and the spin diffusion length of Pt, $\alpha_H \lambda_{pt}$, or equivalently the product of the spin Hall conductivity, electrical resistivity, and the spin diffusion length, $\sigma_H \rho_{pt} \lambda_{pt}$, is used as a figure of merit for the spin Hall efficiency in Pt. We have determined an average value of $\alpha_H \lambda_{pt} = (0.142 \pm 0.040)$ nm for 4 nm Pt at 10K. A gradual decrease of the average $\alpha_H \lambda_{pt}$ at higher Pt thickness is observed. Broad distribution of individual $\alpha_H \lambda_{pt}$ values from 0.03 nm to 0.21 nm is present and indicates possible variations of microstructures. The substantial values of $\alpha_H \lambda_{pt}$ suggest efficient generation of spin current via the spin Hall effects.

## VI, Acknowledgements

We acknowledge use of University of Maryland NanoCenter facilities. Luo, Zhou, and Wu acknowledge the support by National Key Basic Research Program (Grant No. 2015CB921401) and National Science Foundation (Grant No. 11434003, No. 11474066) of China.